\documentclass[smallextended,english]{scrartcl}      
\pdfoutput=1

\usepackage[utf8]{inputenc}

\usepackage{graphicx}
\usepackage{%
    amsmath,
    amssymb,
    babel,
    booktabs, 
    csquotes,
    color, 
    enumitem, 
    geometry, 
    graphicx, 
    lineno,
    multirow,
    natbib,
    siunitx,
    tabularx,
    soul
       }

\usepackage[dvipsnames]{xcolor}

\usepackage{authblk}

\begin{document}
\title{Dynamical anomalies in terrestrial proxies of North Atlantic climate variability during the last 2 ka}

\author[1, 2]{Jasper G. Franke}
\author[1]{Reik V. Donner}

\affil[1]{Potsdam Institute for Climate Impact Research, Potsdam, Germany}
\affil[2]{Humboldt University of Berlin, Berlin, Germany}

\date{}
\maketitle

\begin{abstract}

Recent work has provided ample evidence that nonlinear methods of time series analysis potentially allow for detecting periods of anomalous dynamics in paleoclimate proxy records that are otherwise hidden to classical statistical analysis. Following upon these ideas, in this study we systematically test a set of Late Holocene terrestrial paleoclimate records from Northern Europe for indications of intermittent periods of time-irreversibility during which the data are incompatible with a stationary linear-stochastic process. Our analysis reveals that the onsets of both the Medieval Climate Anomaly and the Little Ice Age, the end of the Roman Warm Period and the Late Antique Little Ice Age have been characterized by such dynamical anomalies. These findings may indicate qualitative changes in the dominant regime of inter-annual climate variability in terms of large-scale atmospheric circulation patterns, ocean-atmosphere interactions and external forcings affecting the climate of the North Atlantic region.

\end{abstract}

\section{Introduction}
The climate during the last two millennia has been generally considered relatively stable as compared to glacial periods and characterized by an absence of strong large-scale rapid climate changes at decadal-to-centennial time-scales \citep{wanner_mid-_2008,ipcc_climate_2013}. This absence of unprecedented climatic shifts has been interpreted as supportive for the development of modern human societies, whereas past periods of unstable climate conditions have been frequently accompanied by cultural decline \citep{coombes2005environmental,marcott2015holocene}. 

Even though rapid climatic shifts of relevant magnitude have been rare, recent studies have provided various examples of episodes with distinct climatic characteristics differing from the long-term normal \citep{mayewski_holocene_2004,crowley_causes_2000,mann_global_2009,miller_abrupt_2012,ipcc_climate_2013}.
The most prominent examples are the Medieval Climate Anomaly (MCA) and the Little Ice Age (LIA), two periods which are believed to have been of substantial importance for the history of especially European societies \citep{fagan_little_2000}.

From a dynamical systems perspective, a common feature of many nonlinear systems is that even slow and minor changes in certain parameters can lead to significant and rapid changes of not only the mean state, but also the qualitative type of dynamics of the system \citep{strogatz2014nonlinear}. Such changes may be of different kinds (e.g.~deterministic or stochastic bifurcations as well as noise-induced shifts between coexisting stable states of the system). Such \emph{dynamical transitions} or \emph{regime shifts} \citep{donges_nonlinear_2011,donges_non-linear_2015,schleussner_indications_2015} are commonly accompanied by transient periods during which the system exhibits anomalous dynamics (i.e.~variability patterns that deviate from what is usually observed during periods of ``normal'' dynamics). Thus, detecting such anomalies from time series can indicate episodes during which the system has evolved far away from equilibrium variability. In such a situation, it is likely that the dynamics of some macroscopic observables (in our context, climate variables) cannot be described by a stationary linear-stochastic process. 

According to these considerations, one possible indicator of the aforementioned transitory periods can be the deviation from a stationary linear-stochastic Gaussian process. One key feature of such processes is (statistical) time-reversibility, i.e.\, that every multi-point statistics is invariant under a change of the direction of time \citep{weiss_time-reversibility_1975}. In this spirit, time intervals during which time-reversibility is temporarily lost can serve as indicators of dynamical anomalies. This can be especially useful in case of rather subtle changes, where it is hard to distinguish between different types of dynamics based on other existing methods, especially such from linear time series analysis. Among other existing approaches to identify signatures of such time-irreversibility, the application of so-called horizontal visibility graphs \citep[HVGs][]{luque_horizontal_2009}) has recently proven to provide a particularly useful tool for the analysis of real-world time series \citep{lacasa_time_2012,donges_testing_2013}. Notably, this method has already been successfully applied to study the transition behavior of the North Atlantic subpolar gyre between the MCA and LIA \citep{schleussner_indications_2015}.

In this study, we apply a similar kind of test for time-irreversibility to an ensemble of terrestrial paleoclimate records spanning the last two millennia. We investigate, if the previously reported dynamical anomalies in the Atlantic ocean circulation across the MCA--LIA transition have been accompanied by similar signatures in atmospheric dynamics. Even more, we identify intermittent time periods during the last 2~ka which possibly separated different dynamical regimes of regional climate variability at interannual time-scales, and examine if the corresponding changes have been consistently recorded in a set of paleoclimate proxies covering vast parts of Northern Europe (Fennoscandia and Iceland). We discuss possible climatic implications arising from a comparison with recent findings from the literature and an assessment of the timing of the observed periods of anomalous dynamics of the considered proxies. Although our method provides new aspects to the discussion of internal variability versus responses to external forcing, an unambiguous attribution of the identified dynamical anomalies to either of these two mechanisms is beyond the scope of the present investigations and remains a subject of future research.

The remainder of this paper is organized as follows. In Section \ref{sec:data}, we present the data used in our analysis. Section \ref{sec:method} briefly describes the method of HVG-based testing for time-irreversibility. In Section \ref{sec:results}, the results of our analysis are presented, which are further discussed in Section \ref{sec:interpretation}. In Section \ref{sec:conclusion}, we make some concluding remarks. A detailed motivation and description of the data analysis methods used in this work, as well as additional results supporting our main findings are provided in the Supplementary Material accompanying this paper.

\section{Data}
\label{sec:data}

\begin{figure}
    \begin{center}
    \includegraphics[width=1.0\textwidth]{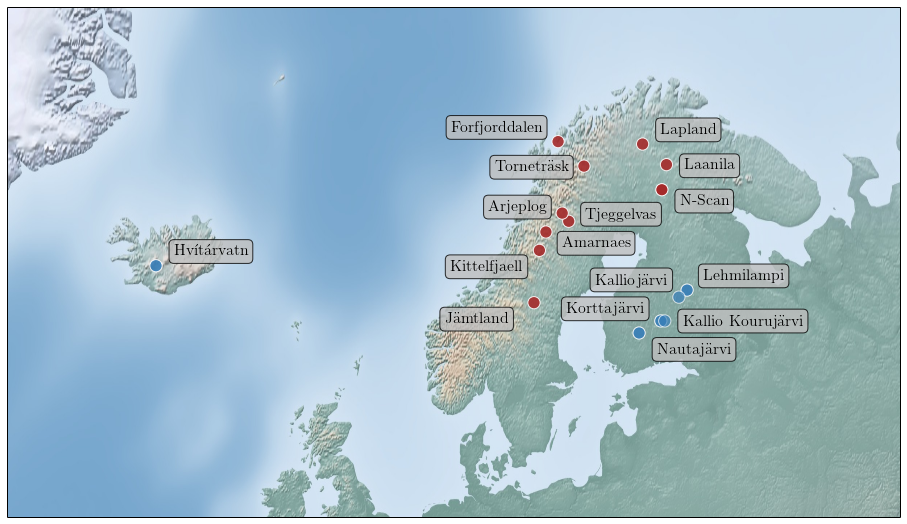}
    \end{center}
    \caption{Geographical distribution of the paleoclimate archives used in this study. Tree ring records are shown in red, lake sediments in blue.}
    \label{fig:map}
\end{figure}

The North Atlantic region is of foremost importance to understand past climate dynamics due to the key role played by phenomena such as the North Atlantic Oscillation (NAO) and Atlantic Multidecadal Oscillation (AMO). It provides numerous very long high-resolution and well-studied tree ring \citep{esper_orbital_2012, melvin_potential_2013,gunnarson_improving_2011,helama_summer_2009} and  lake sediment records \citep{tiljander_3000-year_2008}. Some of these records have been used as key elements in recent regional \citep{pages_2k_consortium_continental-scale_2013,mckay_extended_2014}, hemispheric \citep{ljungqvist_new_2010} and global \citep{mann_global_2009} paleoclimate field reconstructions. 

Individual high-resolution proxy records from the same region commonly differ in their detailed variability due to site-specific effects and archive-specific restrictions to the recording of inter-annual variations. This renders it difficult to obtain a coherent picture of the regional climate dynamics in the past based upon individual records only. In contrast, one needs to consider a paleoclimate proxy as a projection of the regional climate state on a variable of the recording archive (e.g.~tree ring width). This variable is commonly considered to (linearly) depend on one specific climate variable (e.g.~summer temperature). Thus, paleoclimate archives that have been affected by similar regional conditions should exhibit similar dynamics in their proxies, even though the specific variations might differ. Although some of the aforementioned considerations may be questionable, for the remainder of this study, we shall accept them as necessary approximations to the possibly more complex real-world processes. In this work, most of the proxies have been related to (summer) temperatures, even though lake sediment records are influenced by winter conditions as well. All records can be considered interrelated especially on decadal to centennial time scales.

The locations of the records analyzed in this study are displayed in Fig.~\ref{fig:map}, while the (detrended) individual time series are shown in the Supplementary Material Fig.~S1. Further details on the records are provided in the Supplementary Material Tab.~S1. Notably, our selection considers only annually resolved paleoclimate records with at least 300 data points during the last two millennia. 

\section{Methods}\label{sec:method}

In the following, we briefly describe the methodological approach used in this study. More details on all relevant aspects together with the associated background are provided in the Supplementary Material Section~S1.

As mentioned in the introduction, time-reversibility (i.e.\, the invariance of every multi-point statistical characteristic under a reversal of the time axis) is a necessary feature of stationary linear-stochastic Gaussian processes. Thus, detecting any statistically significant violation of time-reversibility for a particular multi-point statistic is a sufficient condition for time-irreversibility and, hence, a violation of the former type of dynamics originating from either system-internal mechanisms or external influences. One example of such a statistic is the asymmetry in the distributions of certain time-ordered characteristics of a HVG, a mathematical transformation from a univariate time series to a complex network representation \citep{luque_horizontal_2009,lacasa_time_2012,donges_testing_2013}. Although not being exhaustive, this approach has the important advantage of being algorithmically simple and computationally cheap. Moreover, the distribution of the associated test statistic (and, thus, the resulting $p$-value) is analytically known.

Specifically, given a time series $\{x_t\}=(x_{t_1},\dots,x_{t_n})$, we identify each individual observation $x_{t_i}$ with the $i$th node of a complex network consisting of $N$ nodes, and pairs of nodes $(i,j)$ with $t_i<t_j$ as being linked if and only if $\min\{x_{t_i},x_{t_j}\}>x_{t_k}$ for all $t_k\in (t_i,t_j)$. The thus constructed set of nodes and links is referred to as the HVG. To construct a statistic sensitive to possible time-irreversibility, for each node $i$, we calculate the fractions of connected triples of nodes $(i,j,k)$ with $t_j,t_k<t_i$ and $t_j,t_k>t_i$ that exhibit all three possible links. These fractions are called the retarded and advanced local clustering coefficients of node $i$, respectively \citep{donges_testing_2013}. Finally, for all observations within a given time window, we compare the empirical distributions of both characteristics using a classical Kolmogorov-Smirnov (KS) test. If these distributions differ significantly (at a confidence level of $\alpha=0.1$), we call the underlying time series in this window HVG time-irreversible, otherwise HVG time-reversible. Here, the relatively moderate choice of the confidence level reflects the possibly short window sizes and the relatively noisy nature of paleoclimate time series.

While the test described above has already been successfully applied in a paleoclimate context \citep{schleussner_indications_2015}, it is known to exhibit a relatively large false positive rate \citep{donges_testing_2013}. Since paleoclimate proxies also vary according to site and archive-specific effects even when recording the same climate signal, we require a significant indication of HVG time-irreversibility to be present for a certain number of subsequent time windows and/or window sizes (see Supplementary Material Section~S1.2). Moreover, since the definition of the HVG is not symmetric under sign-flipping of $\{x_t\}$, we employ the described test for each proxy in parallel to both, $\{x_t\}$ and $\{-x_t\}$ and attribute an indication of time-irreversibility if a corresponding signature is found for at least one of the two possible directions. Finally, since interpreting the obtained significance statements for individual records is not necessarily meaningful \citep{donges_non-linear_2015}, we integrate information from all considered proxy time series by considering only periods during which significant indications of time-irreversibility have been present consistently across the ensemble of records from our study region. For this purpose, we apply a group-wise significance test, which is described in full detail in the Supplementary Material Section~S1.3. The identified intervals with consistent and significant HVG time-irreversibility as discussed below reflect the results of this group-wise test.

\section{Results}
\label{sec:results}
In Fig.~\ref{fig:vg_raw}, the individual $p$-values of the KS tests for all records are shown for the time series $\left\{x_t \right\}$. In Fig.~S2 of the Supplementary Material, the results for one of the time series are displayed in detail as an illustrative example. The complete results for the time series $\left\{-x_t\right\}$ are shown in the Supplementary Material Fig.~S3. We observe that most intervals for which HVG time-reversibility can be rejected at the considered confidence level $\alpha$ have consistently very low $p$-values for most window sizes (blue contours in Fig.~\ref{fig:vg_raw}). The temporal extent of an interval of significant $p$-values might vary, though. By attributing significance statements to intervals instead of individual points in time, we take this time uncertainty into account. In the Supplementary Material Fig.~S1, all detrended time series are shown together with the intervals exhibiting time-irreversibility in each individual record. Some records exhibit considerably more intervals of time-irreversibility than others, which might result from a more non-stationary dynamics. Note that a violation of Gaussianity of the underlying process would not affect the HVG properties, since the latter are independent of the probability distribution of the underlying time series values. Supplementary Material Fig.~S1 clearly highlights the need for the group significance test, as it appears reasonable to assume that some of the identified intervals are false positives.

\begin{figure}
    \includegraphics[width=1.0\textwidth]{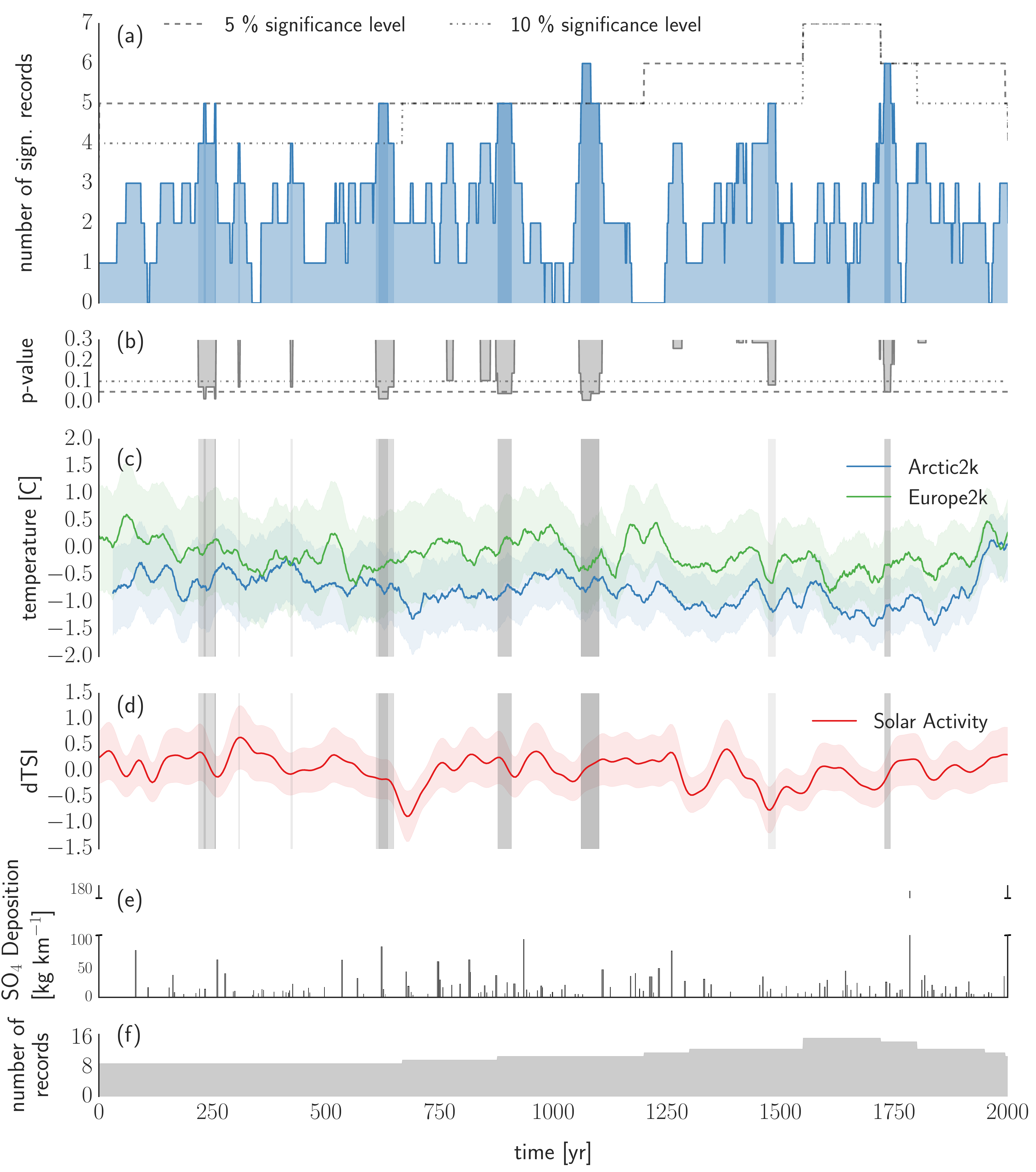}
    \caption{Combined results for the HVG time-irreversibility test: (a) Number of proxies with HVG time-irreversible dynamics per point in time. (b) $p$-value that this number of records show HVG time-irreversibility in the case of underlying stationary linear-stochastic processes. (c) Two regional temperature reconstructions for Europe \citep{luterbacher2016european} and the Arctic \citep{pages_2k_consortium_continental-scale_2013}, respectively. (d,e) State-of-the-art reconstructions of anomalies in the total solar irradiance \citep{steinhilber2009total} and volcanic aerosol deposition \citep{sigl_timing_2015}. Grey vertical bars in panels (b)-(d) highlight the periods of significant HVG time-irreversibility determined by the multi-record significance test. (f) Number of records available at each point in time.}
    \label{fig:p_c_results}
\end{figure}

The results of the multi-record significance test are shown in Fig.~\ref{fig:p_c_results}a,b. We find that there are several episodes throughout the last two millennia that display HVG time-irreversibility in multiple records with different levels of confidence. For comparison, in Fig.~\ref{fig:p_c_results}c we also show two regional summer temperature reconstructions of the past two millennia from \citet{luterbacher2016european} and the \citet{pages_2k_consortium_continental-scale_2013}, respectively. Most of the records considered in this paper have also been used in the Arctic temperature reconstruction \citep{pages_2k_consortium_continental-scale_2013}, however, strong similarities to All-European climate variability are to be expected as well. 

\begin{table}
    \renewcommand{\arraystretch}{2}\addtolength{\tabcolsep}{-1pt}
    \begin{tabularx}{\textwidth}{lXXXX}
        Interval & Time [CE] & $n\ (N)$ & $p$-value & Episode \\
        \toprule
        I1  & 219--257 & 4--5 (8) & 0.016--0.073 &  End of RWP\\
        I2 & 610--649 & 4 (8) & 0.016 &  LALIA/1.4k event\\
        I3 & 878--908 & 5 (11) & 0.041 & Onset of MCA\\
        I4 & 1064--1101 & 6 (11) & 0.009 & Cold episode during MCA/Oort minimum \\
        I5 & 1473--1489 & 5 (13) & 0.082 & Onset of LIA\\
        I6 & 1729--1742 & 6 (15) & 0.049 & Late Maunder Minimum
    \end{tabularx}
    \caption{Time intervals with significant HVG time-irreversibility and coinciding climate phenomena.}
    \label{tab:events}
\end{table}

The identified time intervals with consistent dynamical anomalies are summarized in Tab.~\ref{tab:events} together with their associated $p$-values. Specifically, the following periods are found: 
\begin{figure}
    \includegraphics[width=1.0\textwidth]{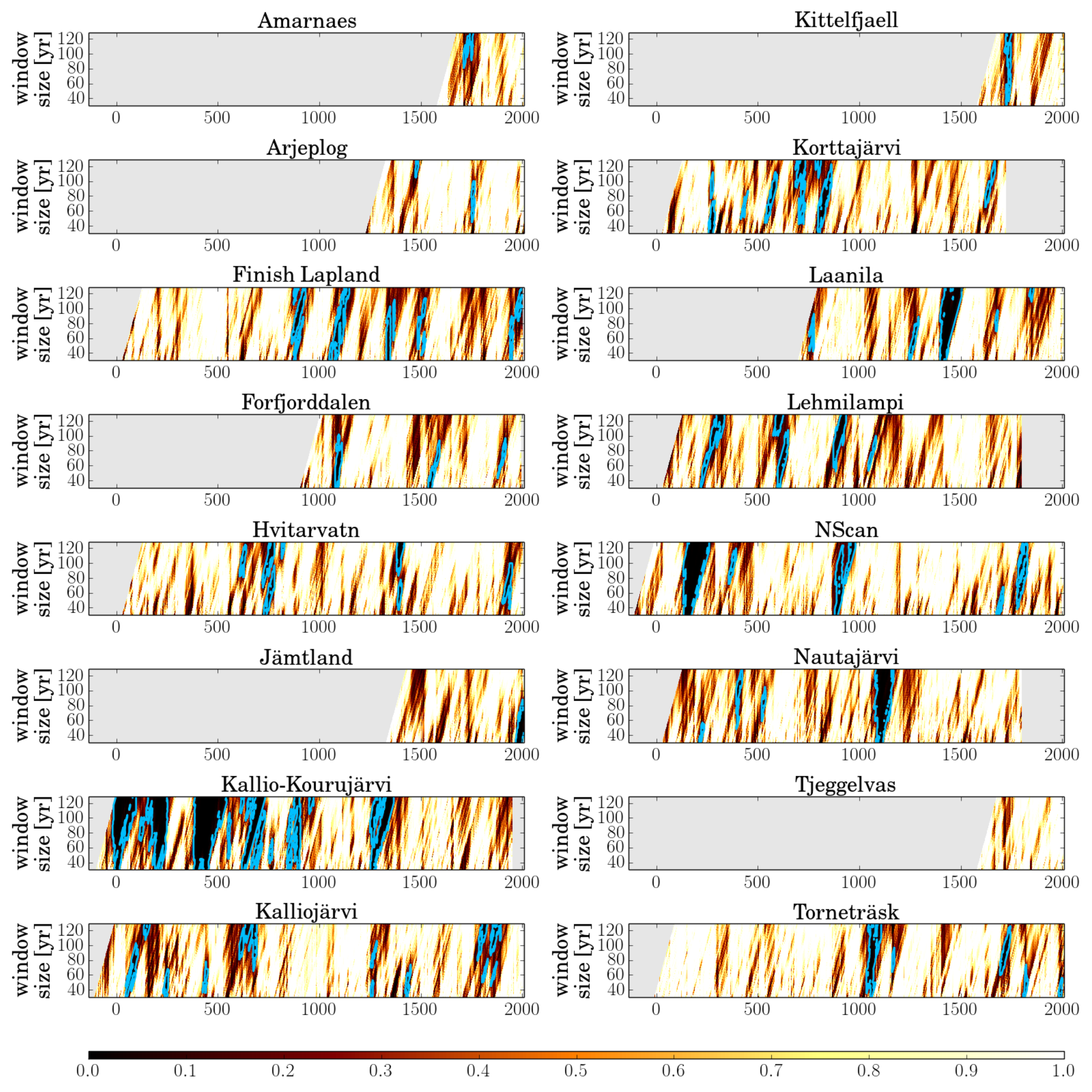}
    \caption{Time-window size planes for all considered records (for $\{x_t\}$). The blue contours highlight areas which are considered to exhibit significant time-irreversibility after sorting out inconsistent signals.}
    \label{fig:vg_raw}
\end{figure}

\begin{itemize}
    \item \emph{Interval I1} precedes a decline of European temperatures starting at about 280 CE and cumulating in a cold period of several centuries duration. This decline is commonly interpreted as the end of the Roman Warm Period (RWP).
    \item \emph{Interval I2} in the 7$^{\text{th}}$ century CE follows the Late Antique Little Ice Age (LALIA) from about 530 CE to 660 CE.        
    \item \emph{Interval I3} coincides with the beginning of the MCA, which is commonly considered to start between 900 CE and 950 CE \citep{ljungqvist_new_2010,ljungqvist_northern_2012,ipcc_climate_2013}.
    \item \emph{Interval I4} corresponds to an episode of colder temperatures during the MCA in the 11$^{\text{th}}$ century. 

\item \emph{Interval I5} precedes the onset of the LIA \citep{ipcc_climate_2013}.

\item \emph{Interval I6} follows the late Maunder minumum (1675--1715)
\end{itemize}

According to the complexity of our analysis procedure (see Supplementary Material Section~S1) and the limited amount of existing high-quality records considered in this study, we have performed several additional tests to ensure the robustness of the results described above (see Supplementary Material Section~S2 for details). These tests demonstrate that the intervals I1, I2 and I4 exhibit robust time-irreversibility. I3 is robust in all cases except for the removal of some records. The intervals I5 and I6 are not robust under all performed tests and thus considered only marginally significant.

\section{Discussion}
\label{sec:interpretation}
In the following, we will discuss possible climatic implications of the episodes of statistically significant HVG time-irreversibility identified in Section~\ref{sec:results} (Tab.~\ref{tab:events}) and how they match with present knowledge of climate variability during the last two millenia. 

During the late Holocene, there have been several episodes of distinct climate variability characteristics, including the MCA and the LIA. Much less discussed, but also of substantial importance are the RWP \citep{buntgen_2500_2011} and the cold period during the Dark Ages nowadays referred to as the LALIA \citep{buntgen2016cooling}). 
While the existence of these climatic episodes is nearly uncontested, there is still no complete consensus on the exact timing, duration and magnitude of the associated changes.
Time-irreversibility in the climate system can have different origins including both, complex ``internal'' dynamics as well as non-stationary external forcings. To take this into account, we also applied the same method of testing for time-irreversibility to a reconstruction of solar activity \citep{steinhilber2009total} (Fig.~\ref{fig:p_c_results}d). The result of the latter analysis is provided in the Supplementary Material Fig.~S4 and reveals three time periods of HVG time-irreversibility in the reconstructed solar forcing, which are compatible with the intervals I2, I5 and I6 during which a multitude of our terrestrial archives showed corresponding dynamical anomalies. A similar analysis for reconstructions of possible volcanic forcing \citep[Fig.~\ref{fig:p_c_results}]{sigl_timing_2015} is hampered by the more intermittent character of eruptive volcanism together with the comparatively lower reliability and representativeness of existing data sets in comparison to solar activity and thus remains as a subject of future work.

\subsection{Roman Warm Period (RWP)}

The RWP is believed to have considerably impacted social and cultural development across Europe \citep{buntgen_2500_2011}. However, little is known so far about its climatic origins. \citet{buntgen_2500_2011} describe increased climate variability in Europe from about 250 to 600 CE and link these conditions to major socio-economic disturbances during this period. Our results indicate that the climatic shift towards a pronounced cold period in the third century CE, which is visible in the regional reconstructions, is also identified as a dynamical anomaly indicating a transition period in the Northern European terrestrial paleoclimate records. This is in good agreement with the literature \citep{bianchi_holocene_1999,pages_2k_consortium_continental-scale_2013}. \citet{luterbacher2016european} found that the first century CE constituted the warmest century of the last two millenia in their reconstruction of European temperatures. Our results are consistent with the hypothesis, that the RWP could have represented a distinct warm state of the regional climate system during the late Holocene, which was terminated by a period of anomalous dynamics finally leading to another more stable climatic situation with generally colder temperatures. 

\subsection{Late Antique Little Ice Age (LALIA)}

Climate variability during the LALIA has recently gained attention due to an increased number and quality of proxies of both climate conditions and possible external forcings. While an extreme volcanic event around 536 CE has long been known \citep[see, e.g.,][]{stothers1984mystery}, the cooler conditions during the following century have just recently been discussed more deeply. \citet{sigl_timing_2015} found a cluster of strong volcanic eruptions at about 536, 540 and 547 CE. \citet{buntgen2016cooling} associated this cluster of volcanic eruptions, in addition to decreasing solar activity \citep{steinhilber2009total}, with a cold period that lasted for about one century until $660$ CE. Thus, recent reconstructions consider the LALIA as an extended period of cooler temperatures, which can also be linked to societal changes and migration in Europe \citep{buntgen2016cooling}. 

In Fig.~\ref{fig:p_c_results}, we can see that our interval I1 coincides with large volcanic eruptions and decreasing solar variability and is followed by a pronounced solar minimum. In the Supplementary Material Fig.~S4, we demonstrate that during this period, solar variability also showed clear signatures of HVG time-irreversibility. Thus, we hypothesize that anomalous variations of solar activity contributed to the complex climatic variability during this time interval.

\subsection{Medieval Climate Anomaly (MCA)}
The MCA corresponds to an episode of relatively warm temperatures mostly observed in Central and Northern Europe, but also visible at a global scale \citep{hughes_was_1994}. The MCA has been related to an extended period of stable atmospheric conditions, dominated by positive phases of the North Atlantic Oscillation (NAO) from the twelfth century onward \citep{bradley_climate_2003,mann_global_2009,trouet_persistent_2009,bradley2016medieval,ortega_model-tested_2015}.

In our analysis, we have identified two periods of complex dynamics associated with the MCA, one around its initiation phase at about 900 CE (I3) and the other from about 1050 to 1100 CE (I4). We interpret the interval I3 as an indicator of a transition period between the colder conditions that dominated the LALIA and a phase of warmer temperatures and hydrological anomalies in the North Atlantic region. Interval I4 corresponds to the recovery phase from temporarily reduced temperatures in the 11$^{\text{th}}$ century leading to a persistent phase of more stable atmospheric conditions favored by the absence of substantial cooling perturbations. 

The cooler period preceding the Medieval Maximum corresponds to a minimum in solar activity \citep{steinhilber2009total} and is a consistent feature in many reconstructions \citep{moberg_highly_2005,ipcc_climate_2013,mann_global_2009,luterbacher2016european} and model simulations \citep{ammann_solar_2007}. This cold period is even more pronounced in some local temperature reconstructions of Fennoscandia \citep{gouirand_spatiotemporal_2008}, which might be explained by a particularly strong local response to solar variability changes \citep{mann_global_2009,bradley_climate_2003}. 
The climatic stability of the MCA might have been supported by an absence of relevant dynamical anomalies in the solar forcing (Supplementary Material Fig.~S4).

\subsection{Little Ice Age (LIA)}
The LIA is commonly described as a sequence of episodes of low temperatures most probably triggered by low solar activity and explosive volcanism \citep{briffa_influence_1998,miller_abrupt_2012,crowley_volcanism_2008,ammann_solar_2007}. 

The interval I5 coincides with the Spörer solar minimum \citep{steinhilber2009total} and two of the largest known volcanic eruptions in 1552 and 1558 \citep{sigl_timing_2015}. \citet{camenisch_early_2016} reported an increased seasonality during the early Spörer minimum and argue, that this was the result of internal variability instead of external forcings alone.

In turn, interval I6 corresponds to the late Maunder minimum, commonly considered the climax and key period of the LIA \citep{luterbacher2001late}. In the latter reference, this pronounced period of cooler temperatures has been attributed to a combination of volcanic and solar forcings and internal dynamics in the North Atlantic.

Both intervals exhibit temporary time-irreversibility in the considered terrestrial proxies, but at a lower confidence level than in most other cases. However, qualitatively this result is in agreement with a similar analysis for two marine records by \citet{schleussner_indications_2015}.

The intervals I5 and I6 coincide with two periods during which the reconstruction of solar activity also exhibited time-irreversibility (Supplementary Material Fig.~S4). While volcanic eruptions have occurred several times during the LIA, they were accompanied by pronounced complex atmospheric dynamics only at the end of the 15th century and during the late Maunder mininum, when solar activity itself also showed time-irreversibility. 

\section{Conclusions}
\label{sec:conclusion}
We have applied the recently developed concept of horizontal visibility graphs (HVG) to detect episodes of time-irreversibility in Late Holocene paleoclimate records from terrestrial archives across Northern Europe.
HVG time-irreversibility indicates that a time series cannot be described by a stationary linear-stochastic process and, thus, that some sort of nonlinear and/or non-stationary dynamics has generated the data under study.

Our analysis identified both the onsets of the Medieval Climate Anomaly and the Little Ice Age as episodes of dynamical anomalies in the recording proxies, which have most likely been triggered by complex, possibly nonlinear atmosphere dynamics at interannual time-scales. In this regard, we provide strong indications for a terrestrial manifestation of the previously reported signature of HVG time-irreversibility in North Atlantic marine records across the Little Ice Age \citep{schleussner_indications_2015}. 

Our observed consistently irreversible time intervals I2, I4, I5 and I6 coincide with periods of low solar activity and often also strong explosive volcanism. Thus, we interpret this time-irreversibility to result not solely from internal dynamics but reflect complex recovery processes after strong external perturbations of the climate system. Regarding time interval I3, effects of external forcing are less obvious given that the MCA has been considered an exceptionally unperturbed period in climate history \citep{bradley2016medieval}. In turn, it is possible that the initiations of both Late Antique Little Ice Age and Littke Ice Age have been rooted in complex variations of solar activity, as for these two periods, we also found time-irreversibility in a reconstruction of solar variability. Thus, our results indicate that the Late Holocene climate dynamics in the European North Atlantic region can be considered as originating from a dynamical system, which has been repeatedly driven out of equilibrium by a combination of reduced solar activity, explosive volcanism and internal feedbacks followed by a complex recovery dynamics. We note, however, that the considered proxies serve as complex filters of climate variability. Therefore, an unambiguous attribution of the observed dynamical anomalies in the proxies to either (linear) responses of the atmospheric circulation to (nonlinear and/or non-stationary) changes in the dynamics of external driving variables (solar activity, eruptive volcanism) or a specific type of nonstationarity (with respect to the considered time scales of interest) of the (nonlinear) internal climate variability itself has been out of reach within the framework of the present study, but will require further research.

From the methodological perspective, we have extended the method of HVG-based testing for time-irreversibility to a regional multi-record analysis. As can be seen in Fig.~\ref{fig:p_c_results}, during most times there have been individual records exhibiting significant HVG time-irreversibility. This indicates a high false positive rate of the employed statistical test design and underlines the need for a multi-record analysis. We have shown that such a group-wise significance test, comprising records from different types of archives and a large region, offers profound and consistent results for regional climate dynamics and highlights known regime shifts in climate variability over the last two millennia. The multi-record approach limits the influence of site and proxy-specific effects and reduces the impact of periods wrongly identified as time-irreversible in individual records. 

A common problem for paleoclimate time series analysis is the relatively low number of available high-quality records covering the same time span, making every analysis vulnerable to statistical artifacts. As more of such records are expected to become available in the near future, the method of HVG-based time-irreversibility testing is expected to yield even more reliable results. In case of a higher number of records, one could also use this method to further infer the spatial characteristics of specific dynamical anomalies.

To conclude, the proposed methodological framework provides additional aspects to the study of past changes in climate variability from a dynamical systems perspective, which complement present-day knowledge on periods where the mean climate deviated strongly from the long-term normal. Our analysis has uncovered periods of transitory behavior in the interannual fluctuations of atmospheric variability as recorded in terrestrial proxies of different types. Although we have restricted our attention to archives with regular annual resolution, the method would also be applicable to other types of archives, including such with irregular sampling \citep{schleussner_indications_2015}. However, in the latter case, additional conceptual problems may arise regarding possible effects of changes in the sampling characteristics, which require additional research. 

\section*{Acknowledgements}
This work has been financially supported by the German Federal Ministry for Education and Research (BMBF) via the BMBF Young Investigators Group "CoSy-CC$^2$ - Complex Systems Approaches to Understanding Causes and Consequences of Past, Present and Future Climate Change" (grant no. 01LN1306A) and the bilateral German-Norwegian project "Nonlinear variability and regime shifts in Late Holocene climate: regional patterns and inter-regional linkages in multi-proxy networks and climate simulations" jointly funded by the German Academic Exchange Service (DAAD project no. 57245873) and the Research Council of Norway. The authors thank Johannes Werner and Saija Saarni for fruitful discussions, and the anonymous reviewers of this manuscript for helpful suggestions to improve its overall presentation. The data used in this study have been kindly provided by PAGES 2k, Saija Saarni and Jan Esper. Calculations have been performed with the help of the Python package \texttt{pyunicorn} \citep{donges_unified_2015}.

\bibliography{literature.bib}

\end{document}